\documentclass[12pt]{iopart}

\usepackage{iopams}
\noappendix
\usepackage{color}
\usepackage{graphicx}
\begin{document}

\title[Systematic design study of all-optical delay line]{Systematic design study of all-optical delay line based on Brillouin scattering enhanced cascade coupled ring resonators}

\author{Myungjun Lee$^1$, Michael E. Gehm$^{1,2}$, and Mark A. Neifeld$^{1,2}$}

\address{$^1$ Department of Electrical Computer Engineering, University of Arizona, Tucson, AZ 85721-0104, USA}
\address{$^2$ College of Optical Sciences, University of Arizona, Tucson, AZ 85721-0104, USA}

\ead{mjlee76@email.arizona.edu}
\begin{abstract}
We present a technique to improve the slow-light performance of a side-coupled spaced sequence of resonators (SCISSOR) combined with a stimulated Brillouin scattering (SBS) gain medium in optical fiber. We evaluate device performance of SCISSOR-only and SCISSOR + SBS systems for different numbers of cascaded resonators from 1 to 70 using two different data fidelity metrics including eye-opening and mutual information. A practical system design is demonstrated by analyzing its performance in terms of fractional delay, power transmission, and data fidelity. We observe that the results from the two metrics are in good agreement. Based on system optimization under practical resource and fidelity constraints, the SCISSOR consisting of 70 cascaded resonators provides a fractional delay of $\sim$ 8 with 22 dB attenuation at a signal bit rate of 10 Gbps. The combined optimal SCISSOR (with 70 resonators) + SBS system provides a improved fractional delay up to $\sim$ 17 with unit power transmission under the same constraints.

\end{abstract}

\vspace{2pc}
\maketitle

\section{Introduction}

 Tunable all-optical delay systems that dynamically manipulate the group velocity of light have received a great deal of attention for optical information processing applications such as data buffering and synchronization. Various slow-light devices, including those based on electromagnetically induced transparency (EIT) in atomic vapor, stimulated Brillouin and Raman scattering (SBS and SRS) in optical fiber, and photonic structures in dielectric material, have been explored as potential realizations of a practical delay system [1--10].

As for on-chip approaches, coupled resonators or photonic crystals are promising techniques that would allow easy integration with other electronics or optical components. Many recent demonstrations of coupled resonator optical waveguides (CROW) and side coupled integrated spaced sequence of resonators (SCISSOR) have been designed and fabricated in compact sizes ($\sim$ 10 $\mu\textrm{m}^2$ ) and with the possibility of dynamic delay control and large delay-bandwidth product [2--4].
 A more recent analysis from Otey $\textit{et al}$ shows that cascaded resonators can even capture light pulses (i.e., stopped light) by completely compressing the system bandwidth and that the captured pulse can then be released \cite{otey}.

 A large fractional delay (equivalent to the delay-bandwidth product) can be achieved by a chain of resonators. Unfortunately, these devices suffer a fundamental trade-off between transmission loss and delay, which potentially limits the use of large numbers of resonators. For example, a CROW consisting of 6 ring resonators demonstrated continuously controllable fractional delay up to 3 at a signal bit rate (BR) of 10 Gbps and a bit error rate (BER) of $10^{-9}$ \cite{mori08}. Its transmission loss, however, is 3 dB (i.e., 0.5 dB/ring) and therefore the use of any additional resonator will increase the BER higher than $10^{-9}$. For comparison, Xia $\textit{et al}$ have demonstrated a chain of 56 cascaded micro-ring resonators in a side-coupled configuration using a silicon-on-insulator waveguide. They achieved a large fractional delay ($\sim$ 5) at a BR of 10 Gbps and BER of $10^{-4}$ \cite{xia}. This high BER is a direct consequence of the 22 dB transmission loss of the device resulting from the use of the large number of rings.

  One possible way of preserving acceptable output signal quality without sacrificing the delay performance is to use a Brillouin amplifier. An SBS gain-based delay system could provide significant signal amplification and its tunable gain bandwidth could be increased up to 25 GHz
  , which allows high speed data transmission \cite{song,Zhu}. In addition, an optimal SBS gain system would provide additional fractional delay of up to 3 \cite{hugo}. Therefore, combining this system with cascaded resonators or other photonic resonance structures seems like a promising method for compensating their respective disadvantages while increasing maximum fractional delay [14--16].

 In general, large slow light delay is accompanied by substantial group velocity dispersion (GVD) that manifests itself as signal distortion. The presence of higher-order GVD terms lead to changes in the pulse shape. Under such a condition, we require a metric to quantitatively measure the output data $\emph{quality}$ along with the $\emph{delay}$.
A common measure of communications performance for the propagation of a pulse train is an $\emph{eye-diagram}$. Eye-diagrams are useful for estimating signal distortion via the maximum eye-opening; and its location represents the delay [16--19]. Neifeld and Lee have presented an alternative metric that uses Shannon information to estimate the information capacity and information delay in the presence of noise \cite{mark,mark1}. In this paper, we utilize these two metrics to evaluate SCISSOR, SBS, and SCISSOR + SBS under practical resource and fidelity constraints. By jointly optimizing the system parameters of the SCISSOR + SBS system, we determine the maximum fidelity-constrained fractional delay at a BR of 10 Gbps.

\section{Data fidelity metric}

The important quantities to consider for evaluating slow-light system performance are the fractional delay and received data fidelity. When a single pulse or a pulse sequence propagates through the dispersive media, it undergoes GVD. There are several metrics including the pulse broadening factor \cite{chin09,SHC2}, amplitude and phase distortion \cite{stenner05}, eye-opening [16--19], and mutual information \cite{mark} that have been introduced to quantify the slow-light performance. In what follows, we consider the eye-opening and information-theoretic metrics.

\subsection{Eye-opening metric }

An eye-diagram is used to visualize the shape of communications waveforms and it is generated by repetitively superimposing subsequent traces of a given data stream over a fixed time interval. The eye-opening (EO) is the maximum difference between the minimum value of ``ones" and the maximum value of ``zeros" at the bit center.
 The data distortion (D) can be quantified using the eye-opening and it is defined as

 \begin{equation}
 D=1- \max(EO).
 \end{equation}

If a pulse sequence passes through a dispersive medium, the output signal could be broadened or distorted, and then D will increase due to the increased intersymbol interference (ISI). Note that distortion has a monotonic relationship with BER and D = 0.35 indicates a corresponding BER $\simeq$ 10$^{-9}$, resulting in reliable communication \cite{Lee2,John}.

An eye-opening based delay can be calculated by the time difference $T_{EO}$ between the input and output eye center defined when the EO is maximal. The fractional eye-opening delay (EOD) is defined as the time delay divided by the input pulsewidth $T_{p}$, that is, EOD = $T_{EO}/T_{p}$.

\subsection{Information theoretic metric}

Information theory was first explored by Shannon and information rate has become a standard method to characterize the quality of a communication channel \cite{sh1,sh2}.
Recently, an information theoretic metric was introduced using the mutual information between the slow light input and output signals, in order to measure the information based delay (ID) and information throughput (IT) \cite{mark}.
The IT-metric in this paper is based on the channel model displayed in figure 1(a). Figure 1(b) shows the examples of 3 bits output signal propagated through an arbitrary slow light channel, which may include effects of delay, distortion, and noise.
 The input X is a binary-valued sequence and it is modulated via on-off keying (OOK). The slow-light delay system is represented by the channel operator H$_S$$_L$, where H$_S$$_L$ could represent any kind of delay device. The mutual information (MI) represents the quantity of transmitted data, and estimates how much input information about X is known when the output Y is observed. Thus, the MI can be defined as $I(X;Y) = H(X)- H(X|Y)$,
where $H(X)$ is the entropy of the discrete input X, representing the a priori uncertainty, and $H(X|Y)$ is the conditional entropy after the output is observed \cite{sh1,sh2}.
We assume that the output signal Y is corrupted by additive white Gaussian noise (AWGN) with zero mean and variance $\sigma^{2}$. We also assume that the elements $x_{i}$ of
a specific n-bit input sequence X are independent and identically-distributed (IID), leading to a prior probability $p(x_{i})$ = (1/2)$^{n}$.
Under these assumptions, the MI can be written as:

\begin{equation}
I(X;Y) = n+ \int\sum_{i=1}^{M}p(x_{i})p(Y|x_{i})\log_{2}\frac{p(Y|x_{i})p(x_{i}) }{\sum_{j=1}^{M} p(x_{j})p(Y|x_{j})  }dY,
\end{equation}
where $\textit{n}$ is the number of input bits, $\textit{M} = 2^\textit{n}$ is the number of possible $\textit{n}$-bit input sequences, and $p(x_{i},Y)$ is the joint probability density function (PDF) of $x_{i}$ and Y.
The integral over Y in equation (2) can be solved by the Monte Carlo simulation with important sampling.
 Here, $p(Y|x_{i})$ is the PDF of Y conditioned on $x_{i}$ that is expressed by the Gaussian PDF:

\begin{equation}
p(Y|x_{i}) \simeq \frac{1}{ (2\pi \sigma^{2}) ^{nL}} \exp(-\frac{1}{2\sigma^{2}}|Y-H_{SL}x_{i}|^{2}),
\end{equation}
where L is the number of simulation samples used to represent a single Gaussian pulse.

Note that the concept of delay is not easily captured within $I(X;Y)$. In order to apply I(X;Y) to the analysis of slow light systems, we impose a window structure, which confines an input pulse sequence within a finite duration window \cite{mark, stenner08}. With this approach, we can compute the MI between $\emph{X and only that part of Y}$ contained within the output window (OW) as a function of window offset \cite{mark, mark1}. Here we use a simple example to describe the IT-metric, let us first consider an ideal distortion free delay device with $\sigma^{2}$ = 0, as shown in figure 2.
The 3 bits of Gaussian pulses with a 50\% return-to-zero (RZ) modulation format serve as an input, and the Gaussian pulse is defined to have field amplitude $E(t)$ = exp$(- (t/T_{HW})^{2})$, where $T_{HW}$ = $T_{b}$/2 is the bit half-width at 1/$e^{2}$ intensity and $T_{b}$ is the bit period. The 50\% RZ modulation denotes that a logical one is represented by a half-bit wide pulse, therefore, $T_{p}$ = $T_{b}$/2. In order to compute the MI for this example of the three bit transmission, we consider all 8 possible states ($\textit{M}$ = 8), as shown in figure 2(a). We assume that the input bit period $T_{b}$ = 100 ps and the value of delay $T_{D}$ = 400 ps. The input signal is fitted within an input window (IW), and then we can compute the MI between input X within the IW and only part of Y contained in the OW for many different OW locations in figures 2(a) and (b).
We observe the values of MI = 3 bit and 1 bit for the two candidate output windows (OW1 and OW2) at two different values of window offset = 400 ps and 600 ps, respectively, as shown in figure 2(b). For this example, when the window offset is the same as the delay $T_{D}$, all the input signal information can be transferred without loss caused by distortion, noise, and energy leaking outside the window.
Thus, the peak value of I(X;Y) represents the amount of information that can be transmitted through the slow light channel; while the location of this peak provides an information-theoretic measure of delay. Therefore, we define the peak-height as the information throughput (IT)  and the peak location as the information delay (ID) of the SL device, where the normalized IT is
\begin{equation}
IT = \frac{\max\{I(X;Y)\} }{\textmd{n-bits} },
\end{equation}
and this definition will be used throughout the remainder of the paper.

\section{Ring resonators}
\subsection{Single resonator}

For a coupled ring resonator, as shown in figure 3, the output fields can be related to the input fields through a complex amplitude transfer function

\begin{equation}
H_{\mathrm{Ring}}(\omega)=\frac{E_{2}(\omega)}{E_{1}(\omega)}=\frac{k-a\ \mathrm{exp}(i\phi(\omega))}{1-ka\ \mathrm{exp}(i\phi(\omega))},
\end{equation}

where $E_{i}$ is the complex field amplitudes, k is the self-coupling coefficient ($k^{2}=1-\rho^{2}$), $\rho$ is a cross-coupling coefficient,
$\textit{a}$ = exp(-$\alpha L_{R}$/2) is the round trip amplitude loss of the resonator, $L_{R}$ is the ring circumference, and $\alpha$ is the total attenuation coefficient which includes all sources of loss such as  material absorption, bending loss, and scattering loss from waveguide roughness \cite{mario,xia}.
 The round-trip phase shift $\phi(\omega)$ in the ring can be represented by $\phi(\omega)$ = 2$\pi n_{R} L_{R}$ $(\omega-\omega_0)$/c, where $n_{R}$ is the effective index of the ring, c is the speed of light, and $\omega_0$ is the resonance angular frequency.

The phase response $\Phi_{\mathrm{Ring}}(\omega)$ of the transfer function is obtained by the relation $H_{\mathrm{Ring}}(\omega)$ = $|H_{\mathrm{Ring}}(\omega)| $exp(j$\Phi_{\mathrm{Ring}}(\omega)$) and it is given in terms of $\textit{k}$, $\textit{a}$, and $\phi(\omega)$ as follows:
 \begin{equation}
\Phi_{\mathrm{Ring}}(\omega) = \pi + \phi(\omega) + \tan^{-1}\Big[\frac{k\ \mathrm{sin}\phi(\omega)}{a-k\ \mathrm{cos}\phi(\omega)}\Big] + \tan^{-1}\Big[\frac{ka\ \mathrm{sin}\phi(\omega)}{1-ka\ \mathrm{cos}\phi(\omega)}\Big].
\end{equation}

Next, we consider the group delay which is a direct consequence of the amount of phase shift in equation (6) within the filter passband. It is defined as the negative derivative of the phase of the transfer function with respect to the angular frequency:

\begin{eqnarray}
 \tau_{\mathrm{Ring}} = -\frac{d\Phi_{\mathrm{Ring}}(\omega)}{d\omega} \nonumber\\
= -\frac{n_{R}L_{R}}{c} + \frac{k(k-a\ \cos\phi(\omega)) }{a^{2}-2ka \ \cos\phi(\omega) + k^{2}}
+  \frac{ka(ka- \cos\phi(\omega)) }{1-2ka\  \cos\phi(\omega) + k^{2}a^{2}}.
\end{eqnarray}

Equations (6) and (7) explain the behavior of the propagated light through the resonator.
At resonance, for $\textit{k} < \textit{a}$, the ring and the waveguide are overcoupled and the phase shift increases rapidly as a function of angular frequency, leading to pulse delay  \cite{heebner,heebner2,blair,Lenz}. On the other hand, for $\textit{k} > \textit{a}$, they are undercoupled and the phase shift decreases rapidly as a function of angular frequency, resulting in pulse advancement. Critical coupling occurs when $\textit{k} = \textit{a}$. Here, the transmission becomes zero at the resonance frequency as the round trip loss of the ring is exactly the same as the fractional loss through the resonance coupling  \cite{O_S}. In our design study, we are particularly interested in pulse delay, and thus all candidate systems use overcoupled resonators.

 In figure 4, we depict the resonator characteristics for four different values of $\textit{k}$ = 0.8, 0.9, 0.95, and 0.97 with a practical value of attenuation coefficient $\alpha$ = 1 cm$^{-1}$ and $L_{R}$ = 150 $\mu$m. Using the numerical simulations based on equations (5) - (7), we calculate and plot the transmission, phase shift, and group delay spectra in figures 4(a), 4(b), and 4(c), respectively. Corresponding Gaussian input and delayed output pulses are shown in figure 4(d), where the input pulsewidth is $T_{p}$ = 50 ps. A silicon waveguide is assumed, thus an effective refractive index $n_{R}$ = 3.0 is used. In figure 4(a), as $\textit{k}$ approaches the critical value of the round trip loss $\textit{a}$ from below, the full width at half depth (FWHD) of the resonator transmission function becomes narrower and deeper. This leads to the slope of the phase shift becoming larger, as shown in figure 4(b), and therefore a larger group delay is achieved for the larger value of $\textit{k}$ = 0.97 in figure 4(c). However, the maximum achievable pulse delay is limited by the tradeoff between the group delay and pulse distortion, causing oscillation at the pulse rising edge, as shown in figure 4(d).

Now, let us consider a pulse train at a BR of 10 Gbps rather than a single pulse, where BR =  1/$T_{b}$.
Figures 5(a), 5(b), and 5(c) present the EO-delay, distortion, and power throughput (PT), respectively, for a resonator as a function of $\textit{k}$ and $L_{R}$. We define the PT as the ratio of the propagated output signal power to the input signal power in the resonator. This numerical simulation is performed by propagating a 127-bit pseudo-random Gaussian pulse train with 50\% RZ modulation format at a BR of 10 Gbps. Our computation covers a range of $\textit{k}$ from 0.94 to 0.99 and $L_{R}$ from 10 $\mu$m to 250 $\mu$m and these ranges are chosen to observe the ring resonator characteristics. It is interesting to note that we observe both the slow and fast light regimes to the left and right sides, respectively, of the critical coupling line (green dashed), in figure 5(a). To increase the delay one can increase $\textit{k}$ or L$_{R}$, but both distortion and energy loss increase at the same time. For a given value of the maximum distortion constraint (e.g. D $\simeq$ 0.35) in figure 5(b), we can find many $\textit{k}$ and $L_{R}$ pairs that provide the same values of distortion-constrained EO-delay $\simeq$ 0.76 with corresponding PT $\simeq$ 0.67, as observed in black dotted lines in figures 5(a) and 5(c). Therefore, we will focus on varying $\textit{k}$ while keeping a fixed practical value of $L_{R}$ = 150 $\mu$m.

\subsection{SCISSOR}
 We now consider a SCISSOR, as shown in figure 6. It is assumed that the SCISSOR has multiple identical rings and its transfer function
$E_{\mathrm{out}}(\omega)/E_{\mathrm{in}}(\omega) =  H_{\mathrm{SCISSOR}}(\omega)=(H_{\mathrm{Ring}}(\omega))^{N}$,
where $\textit{N}$ is the number of resonators. Figure 7 presents the characteristics of the SCISSOR for four different numbers of rings ($\textit{N}$ = 1, 3, 5, and 8) with $L_{R}$ = 150 $\mu$m, $\alpha$ = 1 cm$^{-1}$, and $\textit{k}$ = 0.85. Because the phase shift at resonance for multiple resonators is additive, the magnitude of the total group delay from a summation of the delays of all individual ring resonators increases as a function of $\textit{N}$. The FWHD of SCISSOR transmission resonance also becomes wider than that of a single ring with the resonance transmission
approaching zero. As a result, output pulse power decreases and the output pulse shape becomes more distorted from the its original input shape as SCISSOR length $\textit{N}$ increases, shown in figure 7(d).

\section{Optimal System Design Study}

In this section, we explore optimal system designs for SCISSOR, SBS, and SCISSOR + SBS. Our approach is to maximize the delay performance under practical system resource constraints while maintaining constant data fidelity \cite{MLeeAO, ravi07, ravi08} .

\subsection{SCISSOR}

 We use EO and IT metrics, as described in Section 2, to evaluate the SCISSOR structure. Figure 8 describes the results of the computations summarizing (1) the EO-delay with associated D and (2) the information theoretic delay with associated IT as a function of $\textit{N}$, where three different noise strengths of $\sigma^{2}$= 0.2, 0.3, and 0.4 are used for the IT computation. 
 The EO-based results presented in this paper are based on propagating a 127 bit pseudo-random pulse train with a RZ modulation format at a BR = 10 Gbps. For information-based results, we have utilized 8 bit input sequences with the same modulation format and BR, and therefore a total 256 ( $\textit{M}$ = 2$^{8}$ states in equation (2) ) possible bit patterns are considered. For each input pattern, we use 10$^{6}$ noise samples to obtain reliable results by using a Monte Carlo technique.
   As expected we found that increasing $\textit{N}$ increases both the EOD and ID at the cost of increased distortion. As a result, the normalized IT values decrease. We observe both EOD and ID yield similar delay values, as shown in figure 8(a).
   From figure 8(b), we see that IT decreases faster for higher noise strength with increasing $\textit{N}$, thus the fidelity of information transmission decreases with increasing $\sigma^{2}$ because the decreased signal to noise ratio (SNR) causes information to be lost. Based on D and IT results for the SCISSOR with $\textit{N}$ = 4, distortion of D = 0.342 is measured, while three different values of IT = 0.943, 0.873, and 0.812 are computed with corresponding AWGN levels of $\sigma^{2}$ = 0.2, 0.3, and 0.4, respectively, as shown in figure 8(b). For the given specific noise level of $\sigma^{2}$ = 0.3, one can use at most 4 cascaded resonators while simultaneously maintaining more than 87\% (i.e., IT $\geq$ 0.87 or approximately 7 out of 8 bits) of the transmitted information. Therefore, we take the IT constraint IT $\geq$ 0.87 for $\sigma^{2}$ = 0.3, to correspond with distortion constraint D $\leq$ 0.35.

 Next, under the two signal quality constraints (IT $\geq$ 0.87 and D $\leq$ 0.35), we optimize $\textit{k}$ to maximize EOD and ID for three different attenuation coefficients $\alpha$ = 0, 1, and 3 cm$^{-1}$ as a function of $\textit{N}$ from 1 to 70. The optimal SCISSOR characteristics using both IT and EO metrics are presented in figure 9. We note that the results from the two different metrics provide similar trends.
  When $\textit{N}$ = 70, the maximum fractional delays of approximately 10, 8, and 4 are achieved for $\alpha$ = 0, 1, and 3 cm$^{-1}$, respectively, at the BR = 10 Gbps.
  As $\textit{N}$ increases, both fidelity-constrained-EOD and ID increases, however, early delay saturation for the highest attenuation value is observed in figure 9(a). As $\textit{N}$ increases, $\textit{k}$ must decrease in order to increase the effective FWHD so that the system can satisfy the D and IT constraints. For the same reason, optimal $\textit{k}$ at higher attenuation is smaller than that at smaller attenuation, as shown in figure 9(c). Note that these trends are also explained by the group delay relation of equation (7). The transferred energy in a lossy SCISSOR decreases exponentially because of the induced loss, shown in figure 9(b), and thus, the transmission losses of the $\textit{N}$ = 70 SCISSOR become around 22 dB and 43 dB for $\alpha$ = 1 and 3 cm$^{-1}$, respectively.  This is what would limit the delay performance and reduce the data fidelity of such a system. Inevitably, we must conclude that an amplification process is required for the SCISSOR. As mentioned earlier, an SBS gain medium is a good choice for both increasing the delay performance and the signal amplification by combining it with the SCISSOR.

\subsection{Broadband SBS}

Slow light via the stimulated Brillouin scattering process has previously been demonstrated for tunable delay in optical fiber \cite{stenner05, SHC2,zhu2}. The SBS process is a nonlinear interaction between a strong pump wave and a weak probe wave that is mediated by an acoustic wave. The acoustic wave generated from this interaction scatters photons to the probe wave, shifting the scattered light downward to the Stokes frequency $\omega_{s}=\omega_{p}-\Omega_{B}$, where $\Omega_{B}$ is the Brillouin frequency shift in optical fiber. As a result, the Stokes field experiences strong gain at $\omega_{s}$. For a typical single mode fiber, the Brillouin frequency shift $\Omega_{B}$ is $\sim$ 10 GHz and the Brillouin linewidth $\Gamma$ is $\sim$ 40 MHz near the communication wavelength of 1550 nm. However, this narrow bandwidth limits the achievable data rate to only several megabits per second. Much of the recent research in the SBS slow-light community focuses on broadening the available SBS bandwidth, and several techniques have been experimentally demonstrated that accommodate a GHz data rate. A primary technique is direct modulation of a Gaussian noise source, generated by an arbitrary waveform generator. Gain bandwidths of up to 25 GHz have been experimentally demonstrated \cite{song, hugo, Zhu}.

Under the small signal approximation, the input field $E(0,\omega$) will be amplified at the fiber output according to $E(L_{f},\omega$) = $E(0,\omega)H_{\mathrm{SBS}}(\omega)$, with the SBS transfer function  $H_{\mathrm{SBS}}(\omega)$ = exp($k(\omega)L_{f}$). Here, $L_{f}$ represents the fiber length and $k(\omega$) is the complex wave vector. For the pump broadened SBS, $k(\omega$) = $P_{p}(\omega)\otimes g_{B}(\omega)$ can be obtained by convolving the pump spectrum $P_{p}(\omega)$ with the Lorentzian gain profile $g_{B}(\omega)=g_{0}/ [1-j( (\omega-\omega_\mathrm{s})/(\Gamma/2) )]$, where $g_{0}$ is the line-center gain coefficient. Pant $\textit{et al}$ showed that a super Gaussian function provides a good approximation of the optimal pump profile
$P_{p}(\omega) = (x_{1}/x_{2})$exp$[-( \omega - ((\omega_{s}+\Omega_{B} ))/x_{2})^{2 x_{3}}]$, where the parameters $x_{1}$, $x_{2}$, and $x_{3}$ define pump peak power, pump width, and pump shape (i.e. $x_{3}$=1 is Gaussian and $x_{3} \gg$ 1 becomes nearly rectangular) \cite{ravi08}.  In the next subsection, these three parameters will be optimized subject to the fidelity constraint and the maximum SBS gain constraint, $\textit{G}$ = max\{$k(\omega)L_{f}$\} $\leq$ 10.
When $\omega = \omega_{0}$, the line-center gain of the broadband SBS is defined by $\textit{G} \simeq g_{0} x_{1} \Gamma \pi L_{f}  / 2A x_{2}$, where A is the mode area.
This gain constraint is imposed to avoid the nonlinear amplifier behavior and the maximum available gain $\textit{G}$ = 10 is conservative value as compare to the Brillouin gain threshold of $\sim$ 25 \cite{zhu2}.

\subsection{SBS + SCISSOR}

Recall the results presented in figure 9, from which we proposed the utility of a joint SCISSOR + SBS system. To demonstrate more explicitly the advantages of this combined slow-light device, we present a practical system design, analyzing its performance in terms of several important factors such as FD, PT, D, and IT in this section. The transfer function  for such a device is given by $H(\omega)$ = $H_{\mathrm{SCISSOR}}(\omega) \times H_{\mathrm{SBS}}(\omega)$, and its real and imaginary parts at resonance are related to the gain and the refractive index profiles through the Kramers-Kronig relation. Figure 10 shows the normalized transmission spectra for individual and combined systems along with the spectrum of a 128 bit pseudo-random RZ sequence at BR = 10 Gbps. We assumed the signal carrier frequency and the SCISSOR resonance frequency $\omega_{0}$ are the same as the SBS Stokes frequency $\omega_{s}$.
To better understand the impact of using this combined system, we look at the input and output eye-diagrams after propagating through the combined transmission spectrum, as shown in figure 10, for several different SBS gain values of $\textit{G}$ = 0, 1, 5, and 10. These results are shown in figure 11. For simplicity, we first consider the SCISSOR with $\textit{N}$ = 1 and note that the combined system with SBS $\textit{G}$ = 0 is a resonator-only system.
It is known that the Gaussian pulse propagating through the ring resonator undergoes dispersion effects that can cause oscillations at the pulse rising and/or trailing edges mainly due to the cubic-GVD, as shown in figure 11(b) \cite{mad}. On the other hand, in the SBS system, the output pulse undergoes distortion in the form of pulse broadening mainly due to the quadratic-GVD. Note that distortion management techniques for the SBS system basically suppress the quadratic-GVD term as demonstrated by Stenner \textit{et al} \cite{stenner05}.
By comparing figures 11(b) and 11(d), the fractional EOD for the SCISSOR only and SCISSOR + SBS ($\textit{G}$ = 10) are 0.61 and 1.88 respectively, therefore, it is clearly observed that the SCISSOR + SBS combination not only improves delay performance, but also suppresses the pulse oscillation in the pulse trailing edge arising from the resonator. Although the SBS process also introduces the pulse broadening, it is not significant in this example. In addition, combining SBS + SCISSOR provides additional benefits in terms of delay and PT improvement. Therefore, the combined system provide $\sim$ 3.1 times larger delay with only a small sacrifice of
$\sim$ 1.2 times eye-closing when compared to the SCISSOR-only system. 

 Figure 12 shows the summary of the optimal results for the resonator loss $\textit{a}$ = 1 cm$^{-1}$ as a function of $\textit{N}$ = 1 - 70 for two candidate systems: SCISSOR-only and SCISSOR + SBS systems under data fidelity constraints (IT $\geq$ 0.87 and D $\leq$ 0.35). We observe that results via both metrics agree well. In general, the maximum fidelity-constrained delays gradually increase, while optimal SBS gain $\textit{G}$ and SCISSOR coupling coefficient $\textit{k}$ decrease. The gain and $\textit{k}$ must be chosen effectively to achieve the maximum delay performance under the IT and D limit. Therefore, for the region of N $<$ 7, the maximum gain can remain constant, whereas k decreases. However, any further increase in N requires a decrease in SBS gain as shown in figure 12(c). We know that the presence of loss causes a nonnegligible decrease in PT for increasing $\textit{N}$ as shown in figure 12(b). The results, however, indicate that the combined system can significantly improve the PT and the delay performance. Even for large number of rings ($\textit{N}$ = 70) the combined system can achieve unit power transmission ratio. The optimal design curves presented in figures 12(a) - (d) represent bounds on the performance of our proposed delay devices subject to real-world operating and fidelity constraints.
 In summary, the proposed technique enables a maximum fractional delay of $\sim$ 17, which is $\sim$ 2.1 times the maximum SCISSOR-only delay, with unit power transmission using a cascade of 70 ring resonators combined with an SBS gain medium and can overcome Khurgin's fundamental limit for the fractional delay for the SCISSOR, in which $\textit{N}$ $>$ 100 resonators are required for fractional delay of 10 \cite{khurgin}.

\section{Conclusion}

We have presented a practical system design for increasing the fractional delay while maintaining high data fidelity by combining SBS and SCISSOR. We have employed two different fidelity metrics (EO-metric and IT-metric) to evaluate the slow-light system performance subject to real-world resource constraints. By jointly optimizing the system parameters, the combined SBS + SCISSOR system can provide larger delay and improved power throughput compare to the SCISSOR-only system. We have shown that the maximum fidelity constrained-delay of $\sim$ 17 for SBS + SCISSOR can be achieved with an unit power transmission at a bit rate of 10 Gbps.

\ack
We gratefully acknowledge the financial support of the DARPA/DSO Slow-Light Program.

\appendix
\pagebreak
\Figures
\begin{figure}
\caption{ (a) Channel model of a slow-light delay device. OOK GP generates a train of Gaussian pulses (GP) modulated via on off
keying (OOK). (b) Example of slow-light output pulses including effects of delay, distortion, and noise. Top figures -- without distortion, middle figures -- moderate distortion, and bottom figures -- large distortion causes ISI.}
\end{figure}

\begin{figure}
\caption{Application of the information theoretic analysis to an ideal delay device for the 3 bits transmission (8 possible states). (a) Example signals at the input (dashed) and output (solid) of the slow light operator. The input window (IW) along with two candidate output windows (OW) are shown. (b) Mutual information as a function of output window offset. Note that IT represents the information throughput and ID represents the information delay. }
\end{figure}

\begin{figure}
\caption{System parameters for a single microring resonator.}
\end{figure}

\begin{figure}
\caption{Lossy resonator characteristics for four values of $\textit{k}$ = 0.8, 0.9, 0.95, and 0.97. (a) Transmission spectra, (b) phase spectra, (c) group delay, and (d) Gaussian input and delayed output pulses.   }
\end{figure}

\begin{figure}
\caption{(Color online) Single resonator characteristics. (a) Fractional delay (FD), (b) distortion (D), and (c) power transmission (PT) as a function of $\textit{k}$ and $L_{R}$. The green-dashed line in (a) represents the location of the critical coupling. The black-dotted line in (b) represents D = 0.35, and corresponding distortion-constrained FD and PT are also represented by same black-dotted lines in figures (a) and (c), respectively. SL represents the slow light region and FL represents the fast light region. }
\end{figure}

\begin{figure}
\caption{Side-coupled integral spaced sequence of resonators (SCISSOR).  }
\end{figure}

\begin{figure}
\caption{SCISSOR characteristics for four different number of resonators $\textit{N}$ = 1, 3, 5, and 8. (a) Transmission spectra, (b) phase spectra, (c) group delay, and (d) Gaussian input and delayed output pulses. }
\end{figure}

\begin{figure}
\caption{Summary of EO and IT results for SCISSOR with $\textit{N}$= 1 - 6. (a) EOD and ID for three different values of noise strength. (b) IT on the left axis and distortion on the right axis. The system parameters of $L_{R}$ = 150 $\mu$m and $\textit{k}$ = 0.915 are used.} 
\end{figure}

\begin{figure}
\caption{Optimal results of (a) EO-constrained fractional EOD and IT-constrained fractional ID, (b) power throughput, (c) self-coupling coefficient k on the left axis and $L_{R}$ = 150 $\mu$m on the right axis, and (d) Distortion on the left axis and IT on the right axis as a function of $\textit{N}$ }
\end{figure}

\begin{figure}
\caption{ Transmission spectra (log-scale) for SBS, SCISSOR, and SCISSOR+SBS along with input pulse spectrum at a BR = 10 Gbps. SBS parameters of gain = 10, gain bandwidth = 10 GHz, and super-Gaussian factor = 2 and SCISSOR parameters of $\textit{N}$ = 1, $\textit{k}$ = 0.965, $\alpha$ = 1 cm$^{-1}$, and $L_R$ = 150 $\mu$m are used. }
\end{figure}

\begin{figure}
\caption{Input and output eye-diagrams vs. SBS gain. (a) Input and output eye-diagrams for (b) SCISSOR-only (SBS $\textit{G}$ = 0), (c) SCISSOR + SBS ($\textit{G}$ = 1), (d) SCISSOR + SBS ($\textit{G}$ = 5), and (e) SCISSOR + SBS ($\textit{G}$ = 10). Double-side arrows represent $T_{EO}$. Note that N = 1.}
\end{figure}

\begin{figure}
\caption{Optimal results of (a) EO-constrained fractional EOD and IT-constrained fractional ID, (b) power throughput (dB), (c) SBS gain, and (d) ring parameters $\textit{k}$ on the left axis and $L_R$ on the right axis as a function of $\textit{N}$ = 1-70. Unit transmission is defined as PT = 1 (i.e. PT = 0 dB) }
\end{figure}

\pagebreak

\appendix
\Figures
\begin{figure}[hb]
\centering\includegraphics[width=10cm]{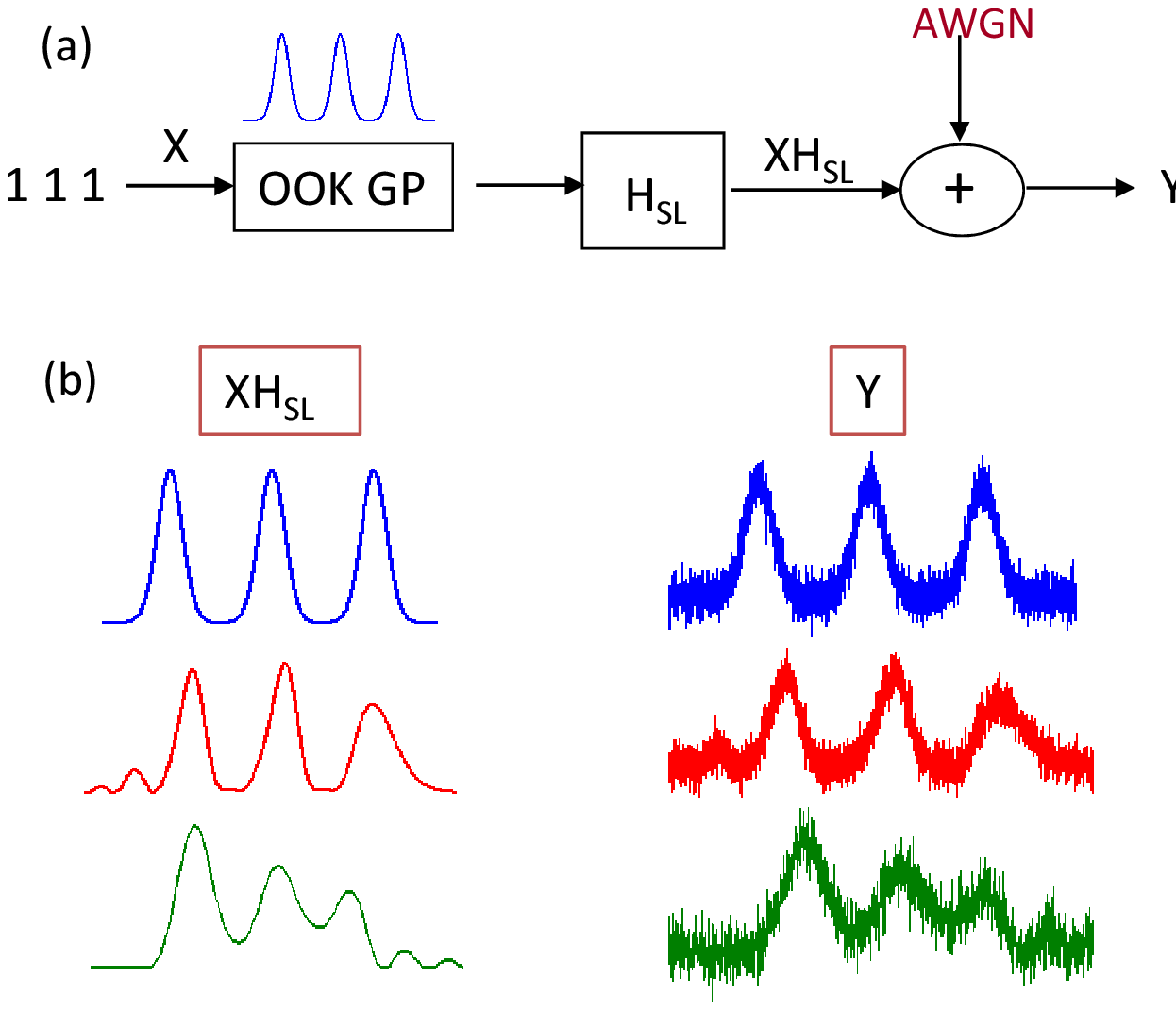}  
\caption{}
\end{figure}
\pagebreak

\begin{figure}[tbh]
\vspace{2in}
\centering\includegraphics[width=10cm]{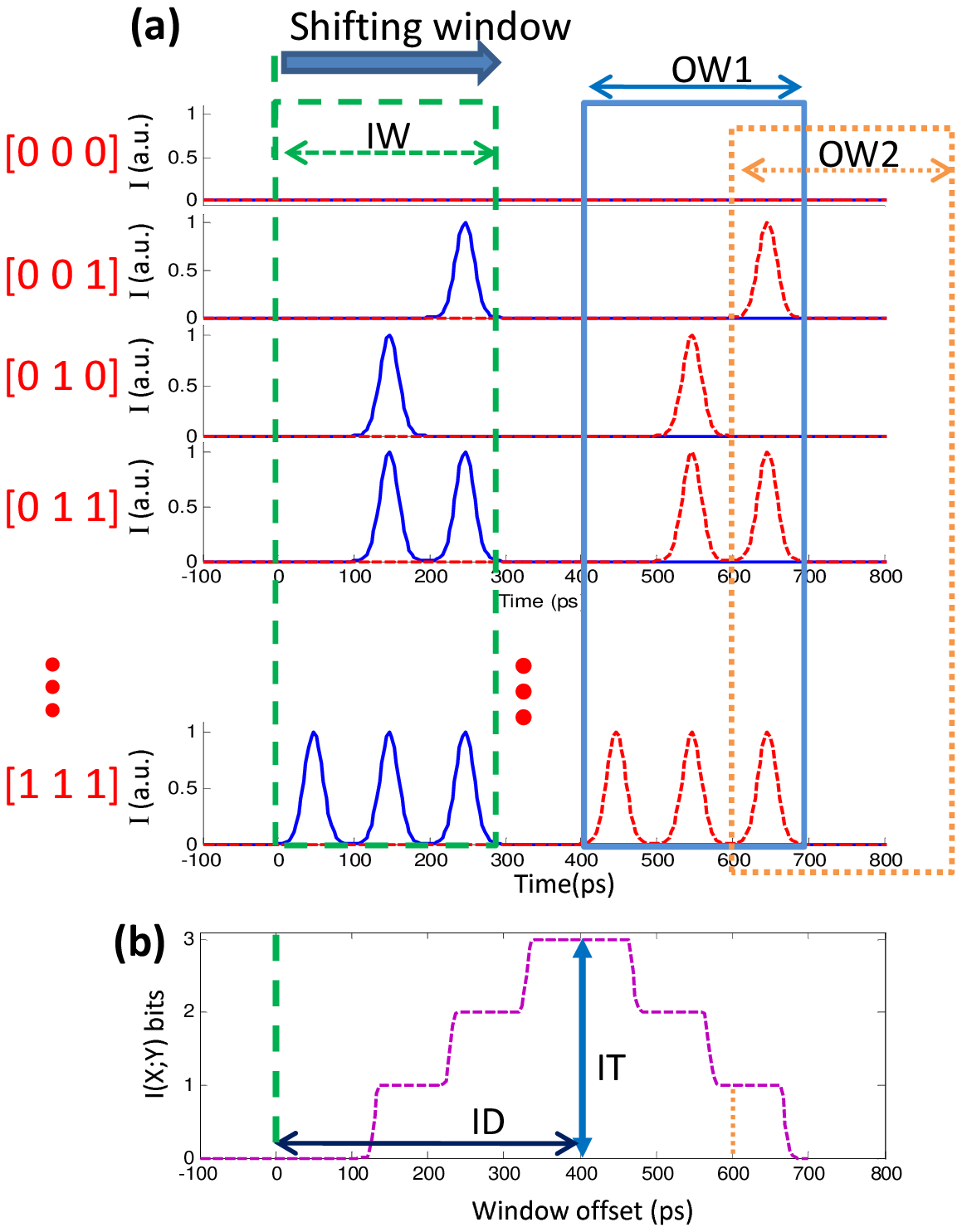}
\caption{}
\end{figure}
\pagebreak

\vspace{2in}
\begin{figure}[b]
\centering\includegraphics[width=5cm]{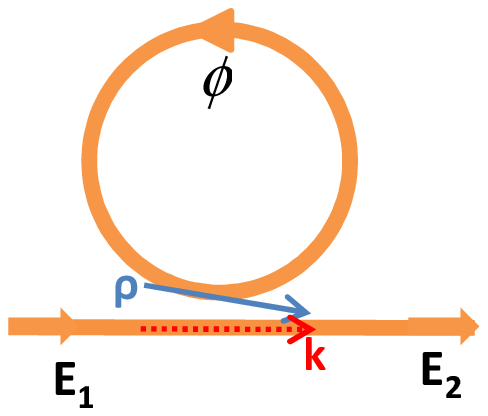}  
\caption{}
\end{figure}
\pagebreak

\vspace{2in}
\begin{figure}[htb]
\centering\includegraphics[width=18cm]{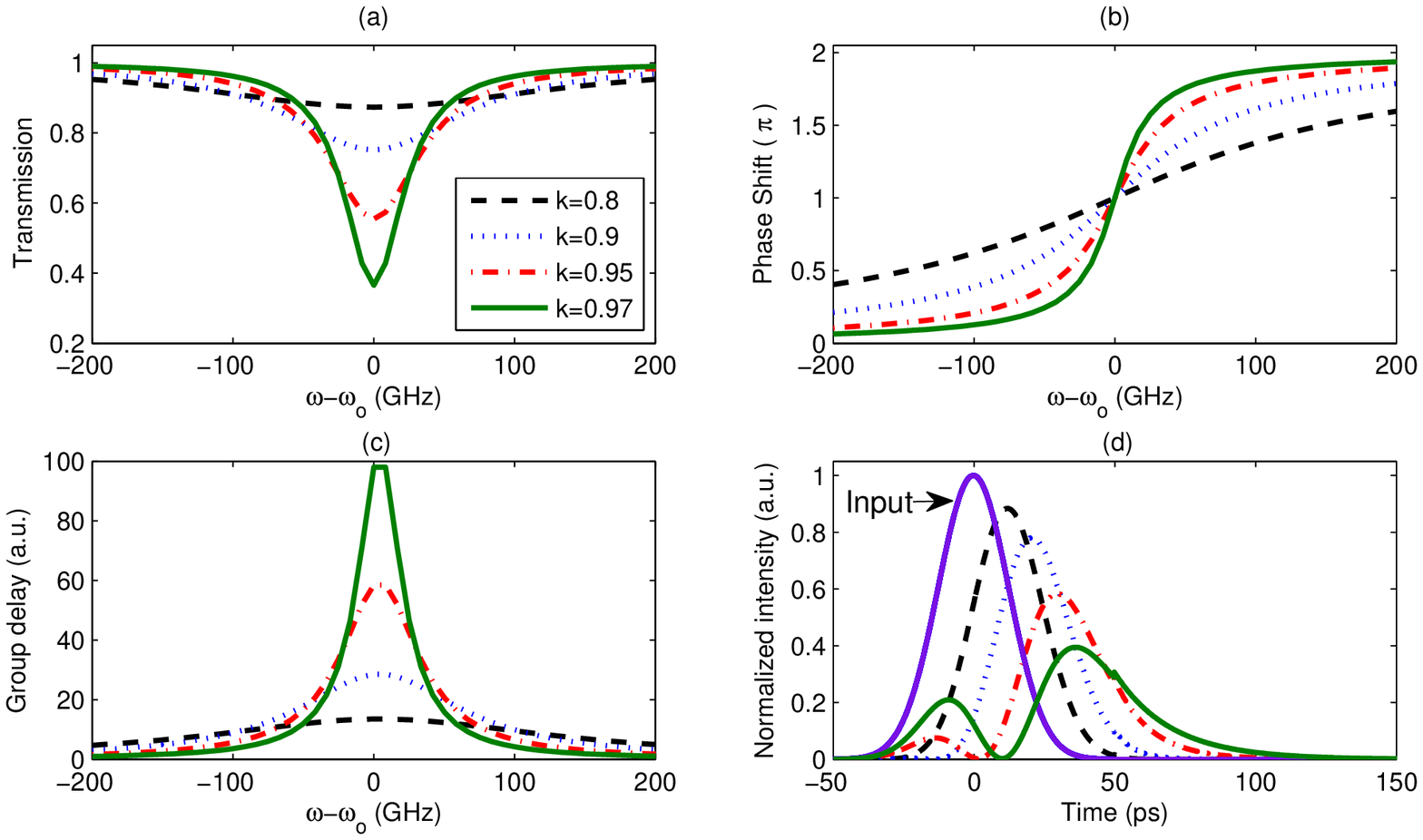}
\caption{}
\end{figure}
\pagebreak

\vspace{3in}
\begin{figure}[htb]
\centering\includegraphics[width=7cm]{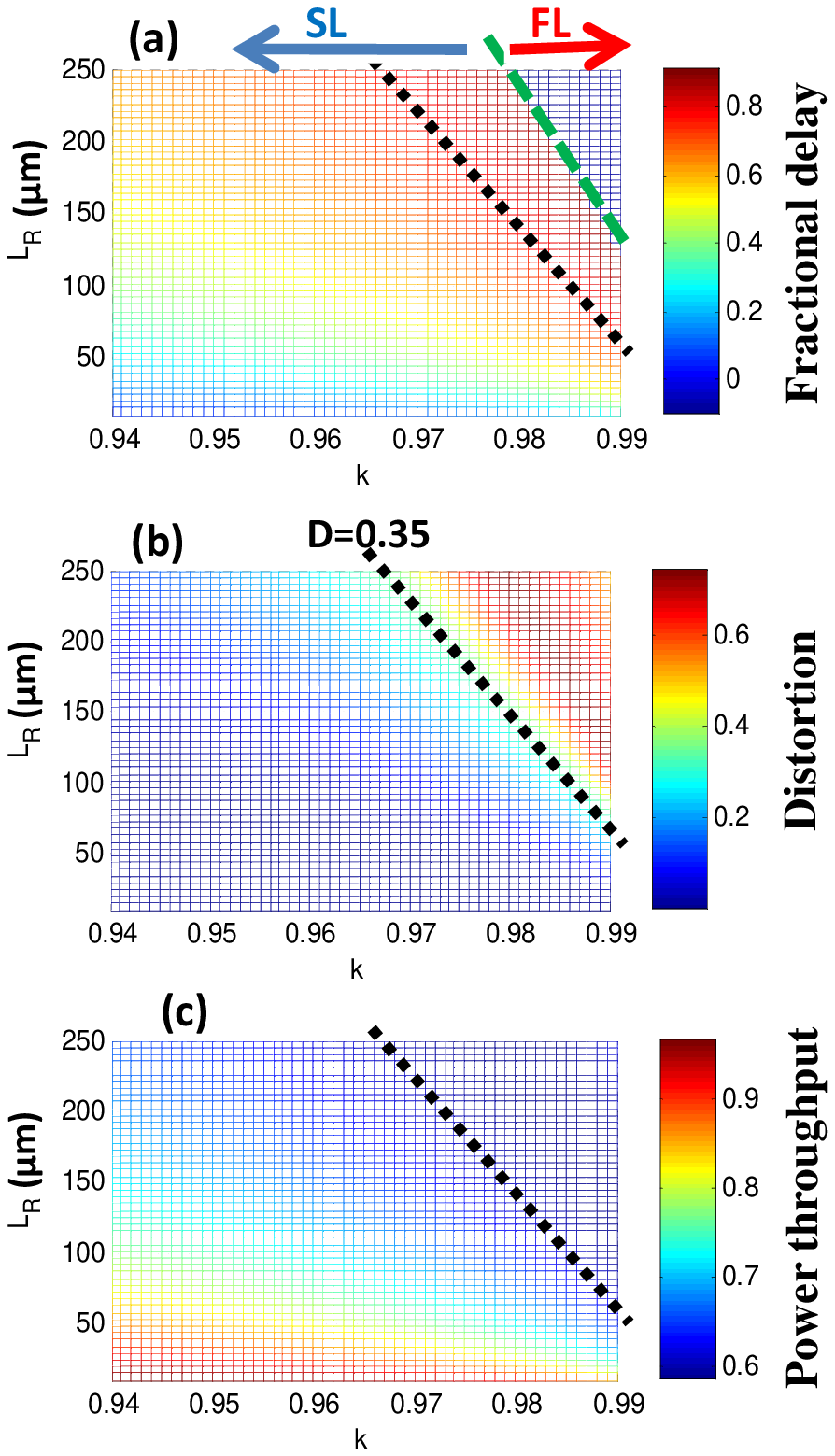}
\caption{}
\end{figure}
\pagebreak

\vspace{1in}
\begin{figure}[hb]
\centering\includegraphics[width=15cm]{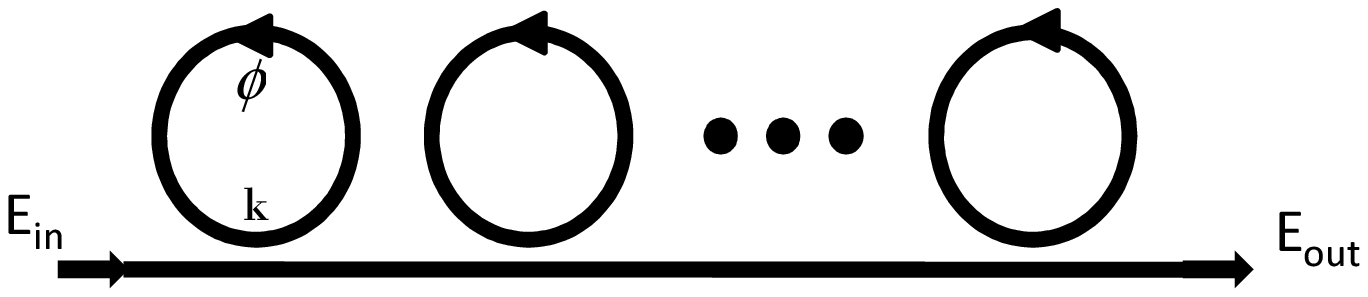}
\caption{}
\end{figure}
\pagebreak

\vspace{1in}
\begin{figure}[htb]
\centering\includegraphics[width=18cm]{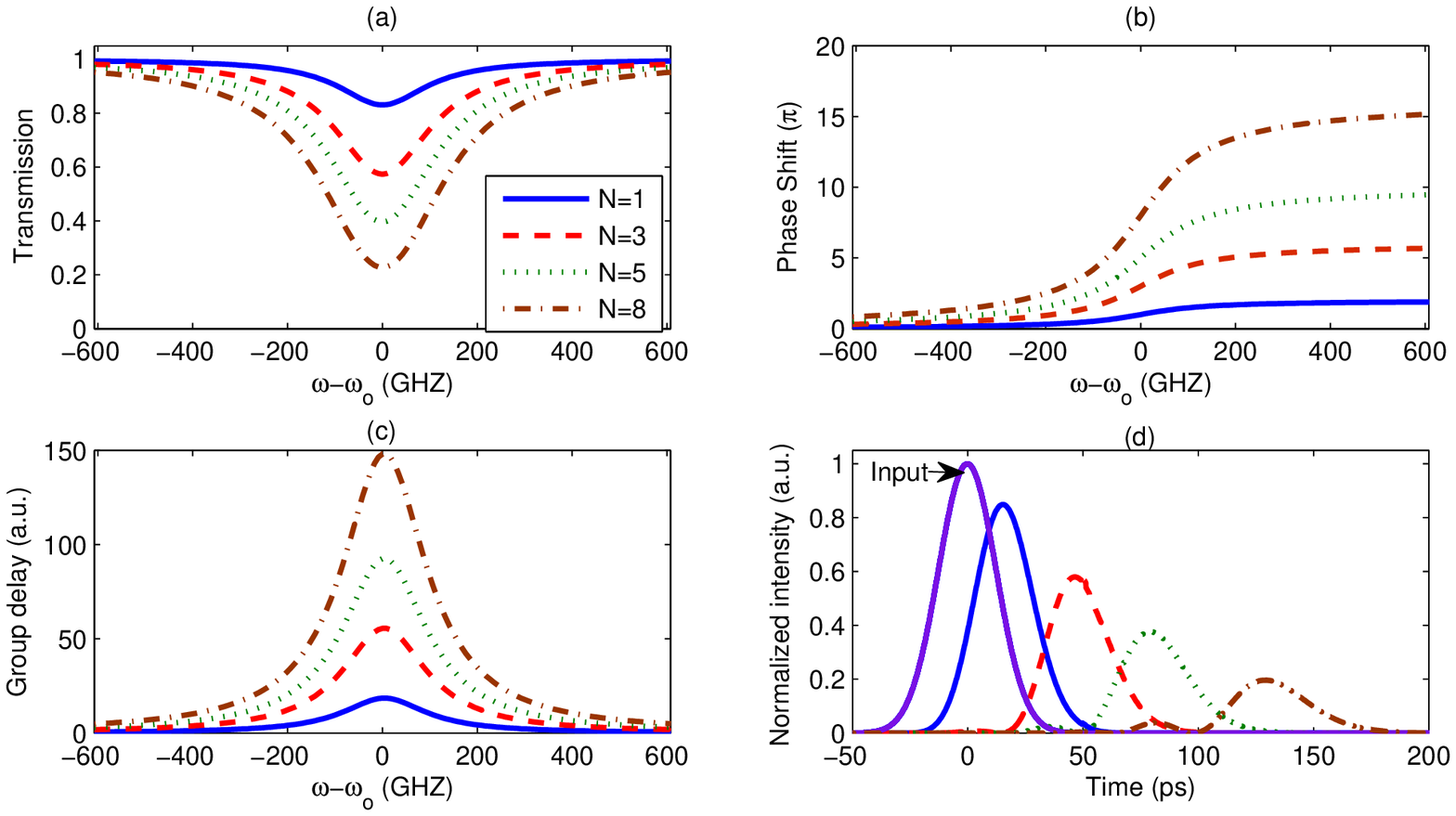} 
\caption{}
\end{figure}
\pagebreak

\vspace{1in}
\begin{figure}[htb]
\centering\includegraphics[width=10cm]{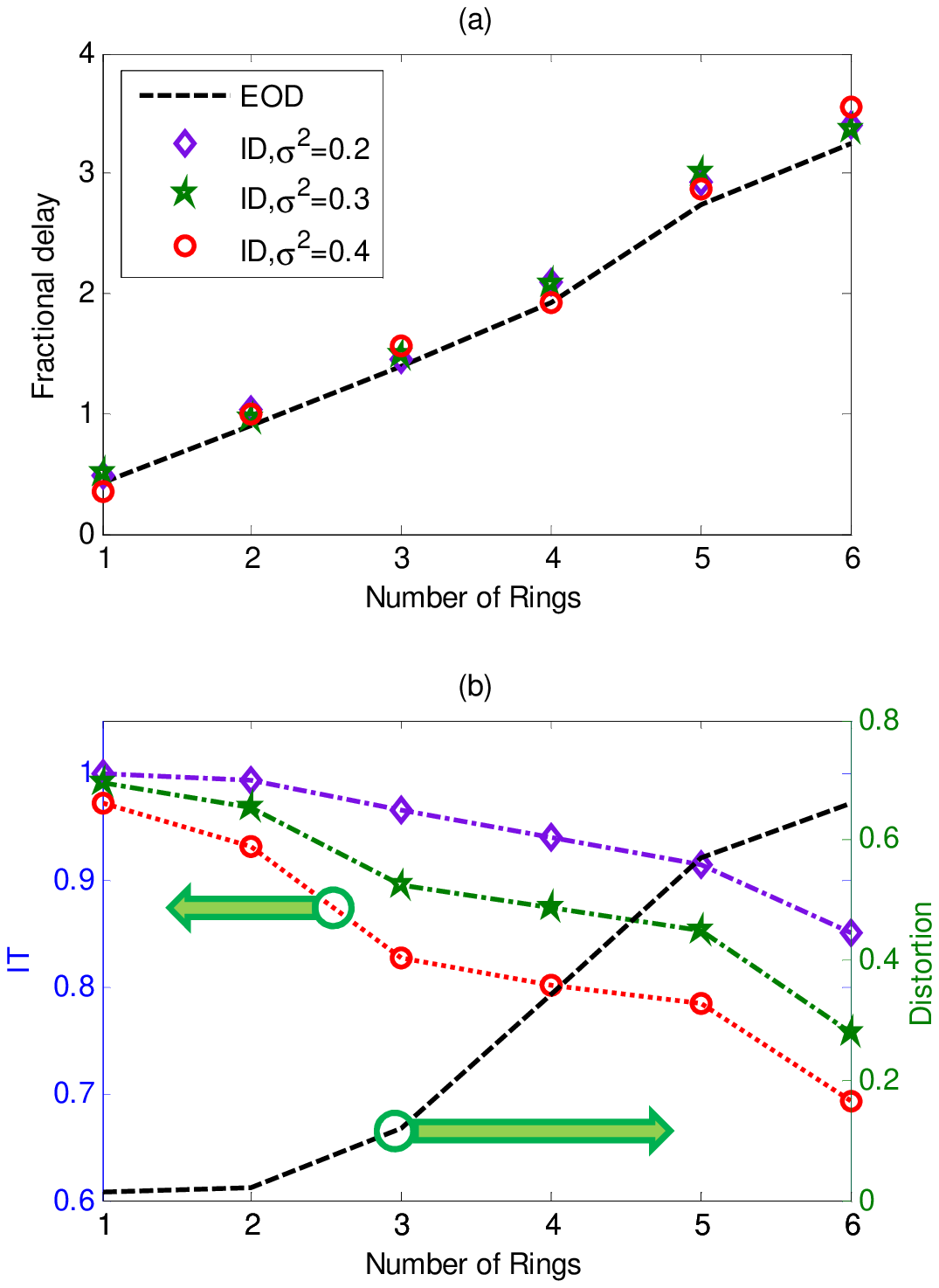}
\caption{}
\end{figure}
\pagebreak

\vspace{2in}
\begin{figure}[htb]
\centering\includegraphics[width=10cm]{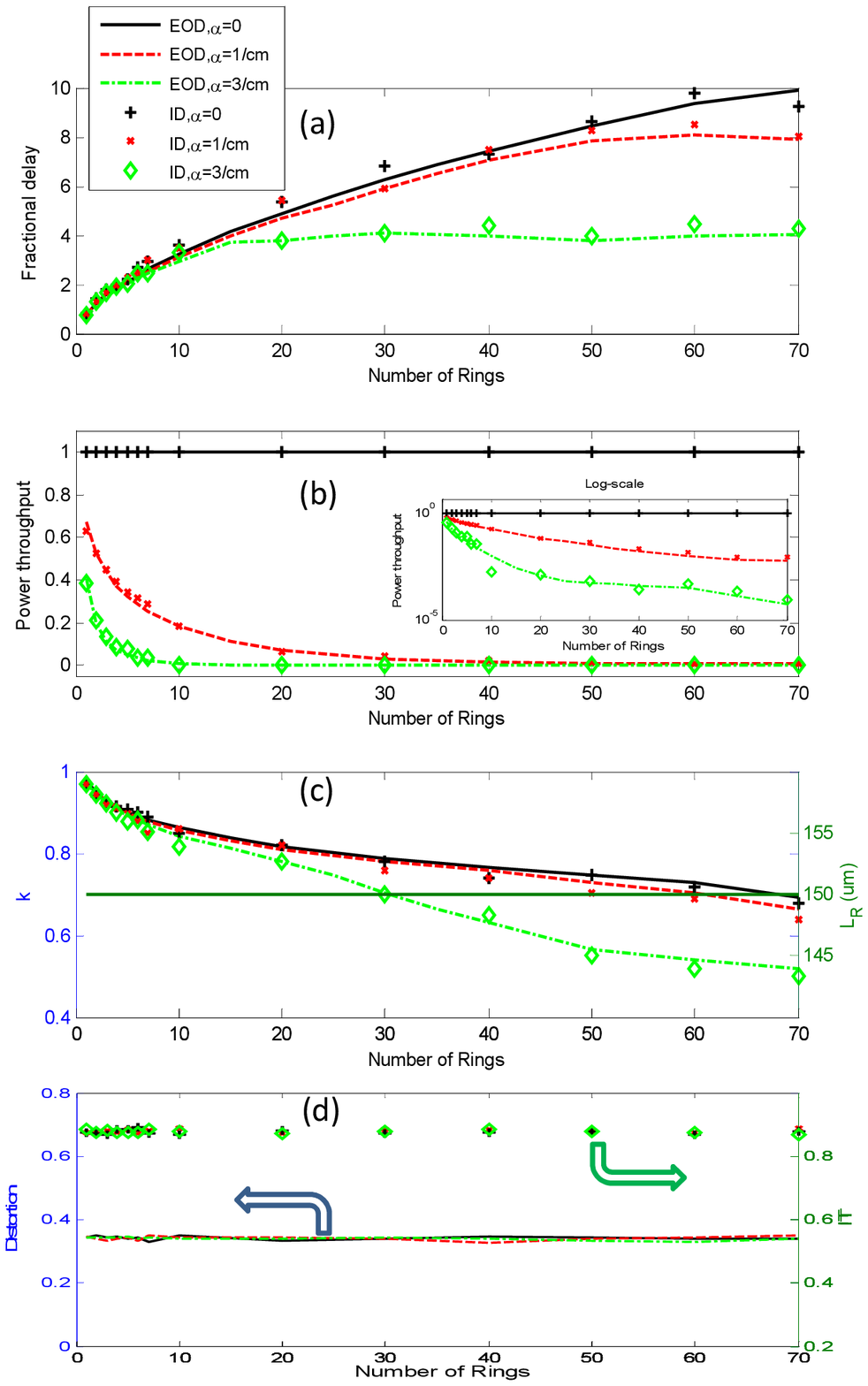}
\caption{}
\end{figure}
\pagebreak

\vspace{1in}
\begin{figure}[htb]
\centering\includegraphics[width=9cm]{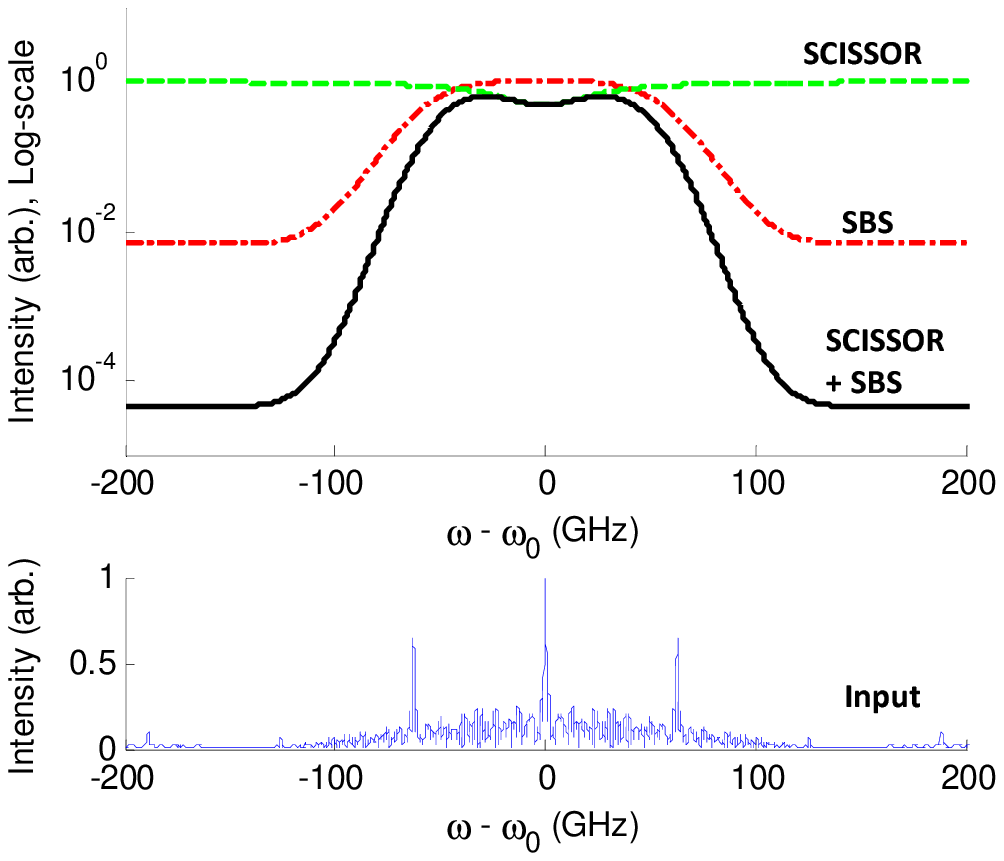}
\caption{}
\end{figure}
\pagebreak

\vspace{1in}
\begin{figure}[htb]
\centering\includegraphics[width=9cm]{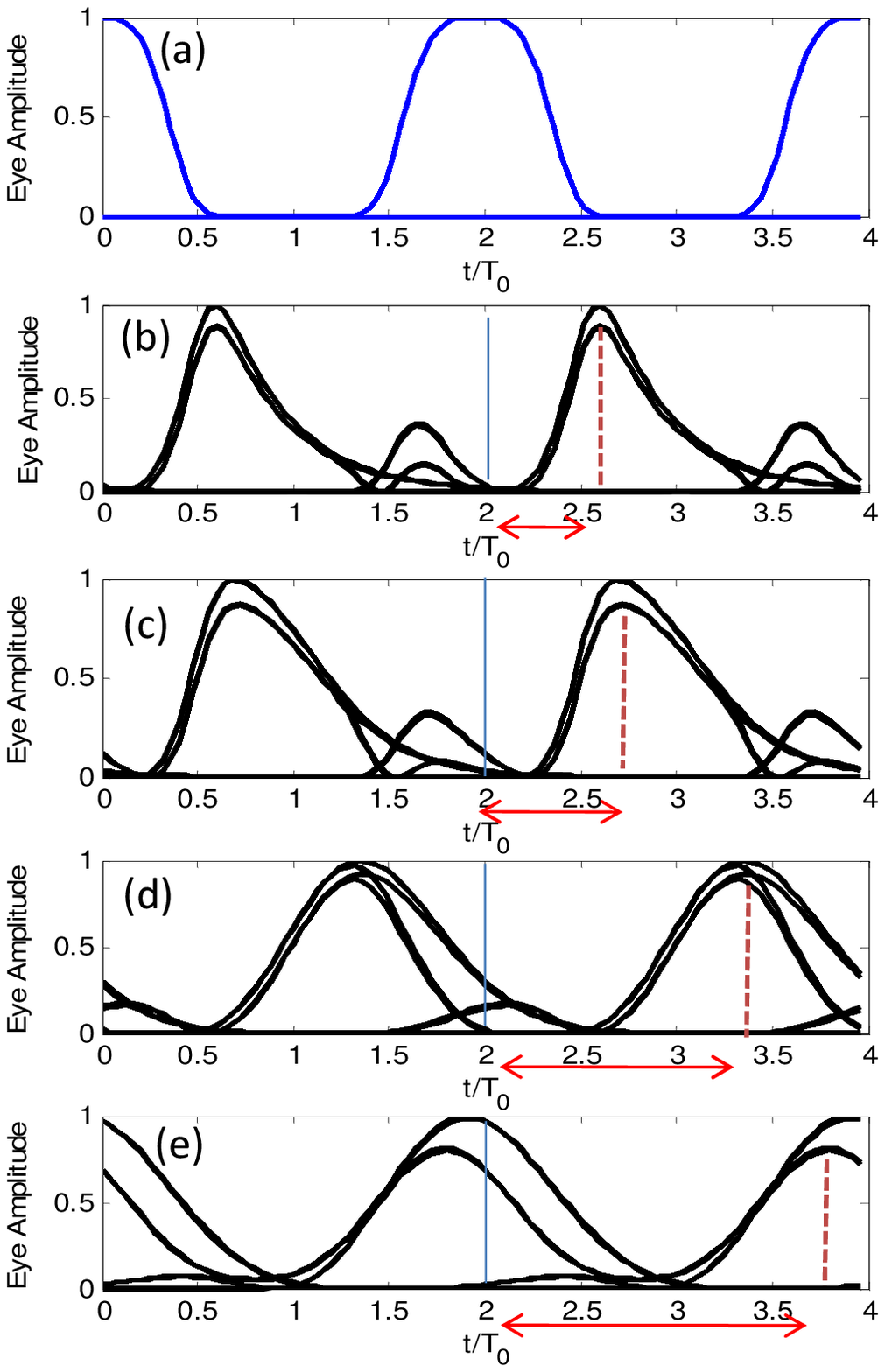}
\caption{}
\end{figure}
\pagebreak

\vspace{3in}
\begin{figure}[htb]
\centering\includegraphics[width=18cm]{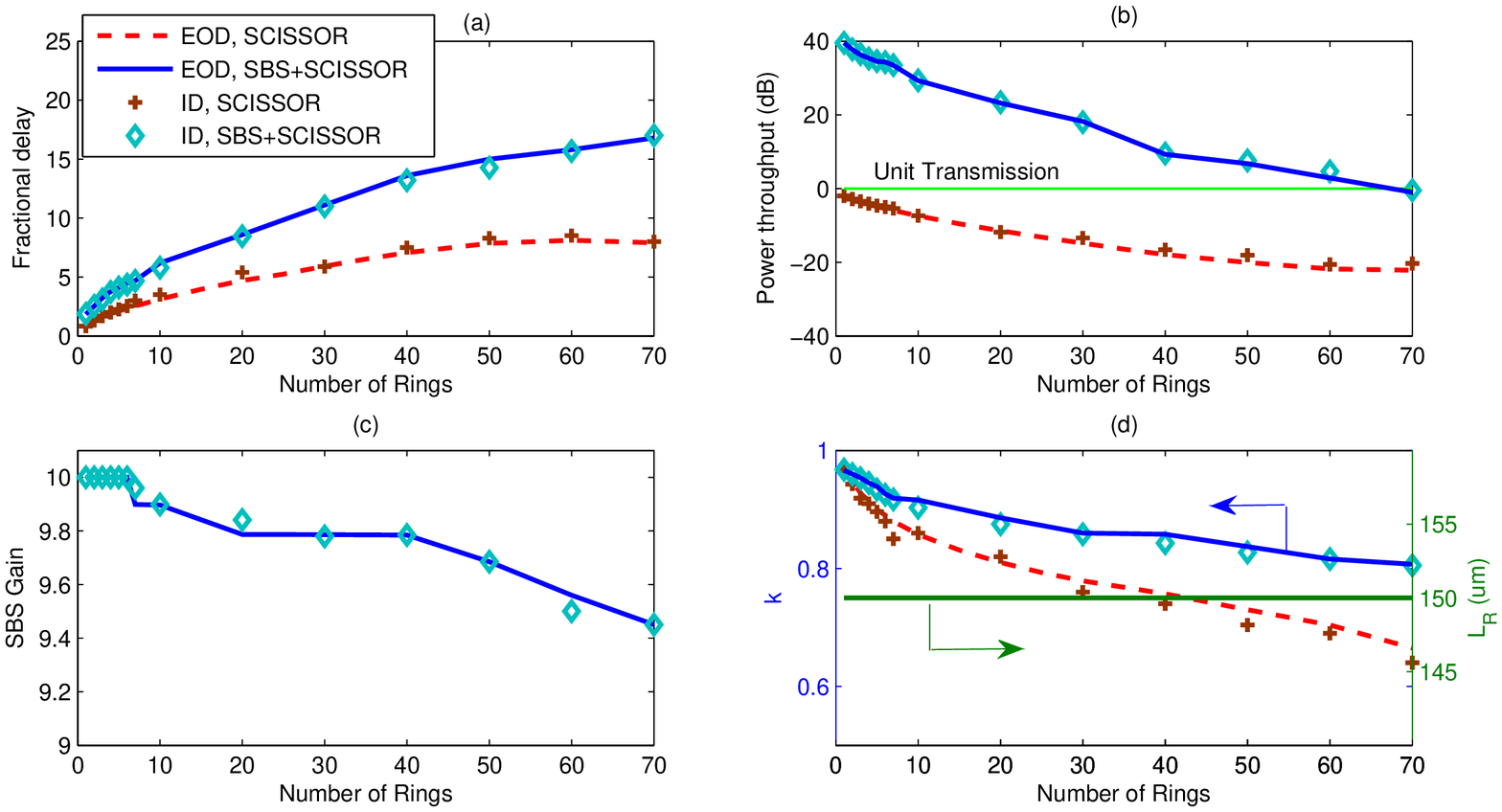}
\caption{}
\end{figure}

\end{document}